\documentclass{pasj00}
\usepackage[dvips]{graphicx}

\begin{document}
\SetRunningHead{Nakagawa et al.}{Astrometry of S~Crt with VERA}
\Received{2000/12/31}
\Accepted{2001/01/01}

\title{VLBI Astrometry of AGB Variables with VERA\\
      -- A Semiregular Variable S~Crateris --}

\author{
Akiharu \textsc{Nakagawa}$^{1}$, 
Miyuki \textsc{Tsushima}$^{2}$, 
Kazuma \textsc{Ando}$^{2}$, 
Takeshi \textsc{Bushimata}$^{3,4}$, \\
Yoon Kyung \textsc{Choi}$^{5}$, 
Tomoya \textsc{Hirota}$^{1,6}$, 
Mareki \textsc{Honma}$^{3,6}$, 
Hiroshi \textsc{Imai}$^{1}$, 
Kenzaburo \textsc{Iwadate}$^{7}$, \\
Takaaki \textsc{Jike}$^{7}$, 
Seiji \textsc{Kameno}$^{1}$, 
Osamu \textsc{Kameya}$^{5,7}$, 
Ryuichi \textsc{Kamohara}$^{3}$, 
Yukitoshi \textsc{Kan-Ya}$^{8}$, \\ 
Noriyuki \textsc{Kawaguchi}$^{3}$, 
Masachika \textsc{Kijima}$^{5}$, 
Mi Kyoung \textsc{Kim}$^{5}$, 
Hideyuki \textsc{Kobayashi}$^{3,4,5,7}$, 
Seisuke \textsc{Kuji}$^{7}$, \\ 
Tomoharu \textsc{Kurayama}$^{3}$, 
Toshihisa \textsc{Maeda}$^{2}$, 
Seiji \textsc{Manabe}$^{6,7}$, 
Kenta \textsc{Maruyama}$^{2}$, 
Makoto \textsc{Matsui}$^{2}$,\\ 
Naoko \textsc{Matsumoto}$^{2}$, 
Takeshi \textsc{Miyaji}$^{3,4}$, 
Takumi \textsc{Nagayama}$^{2}$, 
Kayoko \textsc{Nakamura}$^{2}$, 
Daisuke \textsc{Nyu}$^{2}$, \\ 
Chung Sik \textsc{Oh}$^{3,6}$, 
Toshihiro \textsc{Omodaka}$^{1}$, 
Tomoaki \textsc{Oyama}$^{3}$, 
Nicolas Pradel$^{3}$, 
Satoshi \textsc{Sakai}$^{7}$, \\ 
Tetsuo \textsc{Sasao}$^{9,10}$, 
Katsuhisa \textsc{Sato}$^{7}$, 
Mayumi \textsc{Sato}$^{7}$, 
Katsunori~M. \textsc{Shibata}$^{3,4,6}$, 
Hiroshi \textsc{Suda}$^{7}$, \\ 
Yoshiaki \textsc{Tamura}$^{6,7}$, 
Kousuke \textsc{Ueda}$^{2}$, 
Yuji \textsc{Ueno}$^{7}$, 
and 
Kazuyoshi \textsc{Yamashita}$^{6}$ 
}
\affil{$^{1}$Faculty of Science, Kagoshima University, 
1-21-35 Korimoto, Kagoshima, Kagoshima 890-0065}
\affil{$^{2}$Graduate School of Science and Engineering, Kagoshima University, \\
1-21-35 Korimoto, Kagoshima, Kagoshima 890-0065}
\affil{$^{3}$Mizusawa VERA Observatory, National Astronomical Observatory of Japan, \\
2-21-1 Osawa, Mitaka, Tokyo 181-8588}
\affil{$^{4}$Space VLBI Project, National Astronomical Observatory of Japan, 
2-21-1 Osawa, Mitaka, Tokyo 181-8588}
\affil{$^{5}$Department of Astronomy, Graduate School of Science, The University of Tokyo, \\
7-3-1 Hongo, Bunkyo-ku, Tokyo 113-0033}
\affil{$^{6}$Department of Astronomical Sciences, Graduate University for Advanced Studies, \\
2-21-1 Osawa, Mitaka, Tokyo 181-8588}
\affil{$^{7}$Mizusawa VERA Observatory, National Astronomical Observatory of Japan, \\
2-12 Hoshi-ga-oka, Mizusawa-ku, Oshu-shi, Iwate 023-0861}
\affil{$^{8}$Department of Astronomy, Yonsei University, \\
134 Shinchong-dong, Seodaemun-gu, Seoul 120-749, Republic of Korea}
\affil{$^{9}$Department of Space Survey and Information Technology, Ajou University, 
Suwon 443-749, Republic of Korea}
\affil{$^{10}$Korean VLBI Network, Korea Astronomy and Space Science Institute, \\
P.O.Box 88, Yonsei University, 134 Shinchon-dong, Seodaemun-gu, Seoul 120-749, Republic of Korea}
\email{nakagawa@astro.sci.kagoshima-u.ac.jp}

\maketitle

\begin{abstract}
We present a distance measurement for the semiregular variable 
S~Crateris (S~Crt) based on its annual parallax. 
With the unique dual beam system of 
the VLBI Exploration for Radio Astrometry (VERA) telescopes, 
we measured the absolute proper motion of a water maser spot 
associated with S~Crt, referred to the quasar J1147$-$0724 
located at an angular separation of 1.23$^{\circ}$.
In observations spanning nearly two years, we have  
detected the maser spot at the LSR velocity of 34.7\,km\,s$^{-1}$, 
for which we measured the annual parallax of 
2.33$\pm$0.13\,mas corresponding to a distance of 430$^{+25}_{-23}$\,pc. 
This measurement has
an accuracy one order of magnitude better than 
the parallax measurements of HIPPARCOS. 
The angular distribution and three-dimensional velocity 
field of maser spots indicate a bipolar outflow with 
the flow axis along northeast-southwest direction. 
Using the distance and photospheric temperature, 
we estimate the stellar radius of S~Crt  
and compare it with those of Mira variables. 
\end{abstract} 

\section{Introduction} 
Very Long Baseline Interferometry (VLBI) is a powerful technique 
for obtaining positions of celestial objects with 
milliarcsecond\ (mas) level accuracy. 
The VLBI Exploration of Radio Astrometry (VERA) telescopes are
a Japanese VLBI array dedicated to phase referencing 
VLBI~\citep{kob03}. 
VERA consists of four 20\,m diameter antennas  at Mizusawa, 
Ogasawara, Iriki and Ishigaki-jima (see figure~1 of \cite{pet07}). 
To overcome phase fluctuations and the limited integration time in 
conventional fast-switching VLBI, VERA has a dual 
beam system that allows simultaneous observations of the target and 
reference sources separated by 0.3 to 2.2 degrees. 
This advanced capability of VERA can be used to measure 
annual parallaxes and proper motions of masers 
with a 10 micro arcsecond ($\mu$as) level accuracy.

\begin{figure*}[htpb]
\begin{center}
 \includegraphics[width=130mm, angle=0]{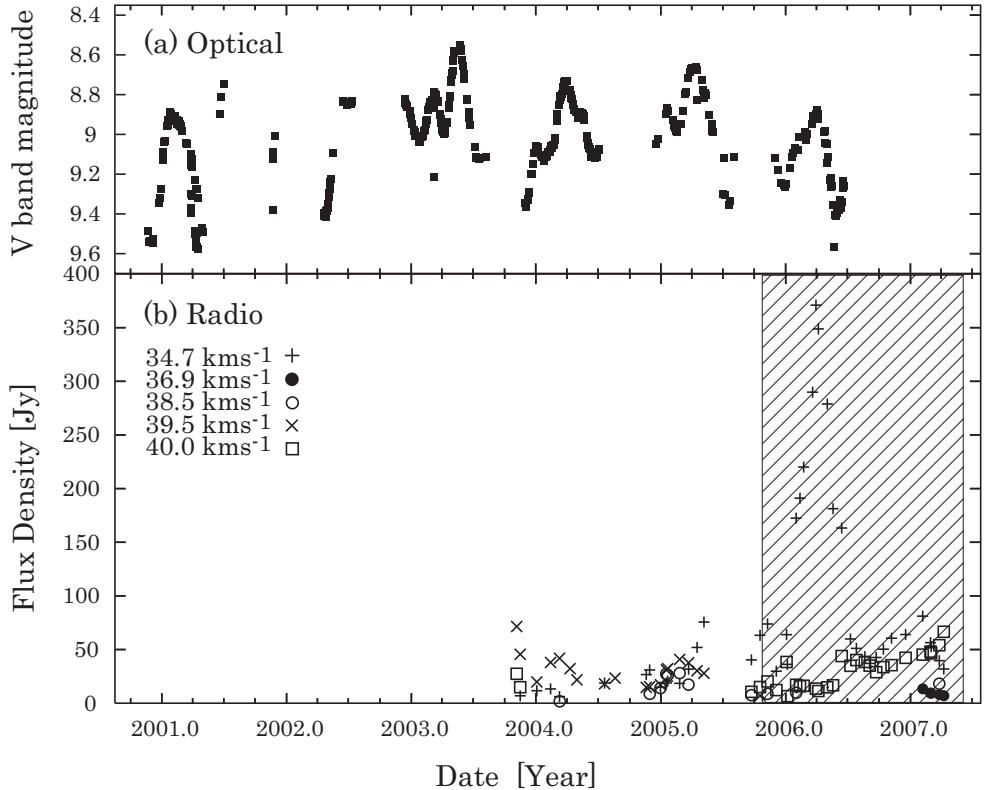}
 \caption{Light curves of S~Crt.
(a)\,The optical light curve at $V$ band provided by the ASAS \citep{poj97}. 
(b)\,The water maser light curve at 22\,GHz obtained with the 
single-dish monitoring program with VERA~\citep{shi08}. 
Peak flux densities 
of major velocity components are presented with different 
symbols. The symbols of ``$+$'', ``{\Large$\bullet$}'', 
``{\Large$\circ$}'', ``$\times$'', and ``$\square$'' indicate 
the peak fluxes of the individual components at the LSR velocities of 
34.7, 36.9, 38.5, 39.5, and 40.0\,km\,s$^{-1}$, respectively. 
There were no available data before the year of 2003.8. The hatched area 
indicates the period of the present VLBI monitoring observations. }
 \label{lightcurves}
\end{center}  
\end{figure*}

An annual parallax gives a simple geometrical measure of the distance, 
free from the complex assumptions required in other distance estimators. 
Studies of Asymptotic Giant Branch (AGB) stars by \citet{whi08}, 
\citet{kur05}, and \citet{vle03} are based on the annual parallax 
derived from VLBI observations, and demonstrated the the ability of 
VLBI astrometry. 
More recently, successful measurements of distances and proper 
motions with VERA have been reported
~\citep{hir07,hon07,ima07,hir08,sat07}. 
In the present paper, we report the observations of 
S~Crt, concentrating on the distance measurement. 
S~Crt (IRAS11501$-$0719, IRC$-$10259, AFGL4830S) is an AGB star, 
with a pulsation period 
of 155\,days\,\citep{ben90}.  
S~Crt has a variability type of SRb in the General Catalog of 
Variable Stars~\citep{ant05}. The apparent magnitude at $K$ band 
is 0.786$\pm$0.314\,mag \citep{cut03}. The optical light 
curve in $V$ band is presented in figure~\ref{lightcurves}(a) 
exhibiting $\sim$0.8\,mag variation. 

The distance to S~Crt has been measured in various ways.
\citet{bow94} and \citet{pat92} determined the distance to be 420\,pc 
and 285\,pc, respectively, using the period-luminosity\,(PL) 
relation of Mira variables. 
\citet{per97} used HIPPARCOS data to measure a parallax of 
2.04$\pm$1.31\,mas, corresponding to a distance of 490\,pc, but with an 
large uncertainty. 
Recently, the new HIPPARCOS catalog~\citep{van07} gives the parallax of 
1.27$\pm$0.92\,mas, corresponding to a distance of 787\,pc. 
VERA offers the potential to determine 
the parallax with a better accuracy so that 
the inconsistencies in these distance measurements can be resolved. 

The observations and data reduction are described in section~2. 
In section~3, we present absolute positions and internal motions 
of maser spots. The annual parallax of S~Crt is presented and 
converted to the absolute distance in this section. 
Finally, in section~4, we examine the properties of S~Crt 
in the context of the new distance measurement. 

\section{Observations and data reduction}
\subsection{Observations}
We have conducted a series of monthly VLBI observations from 
October 2005 to May 2007 with VERA. The duration of each observation 
was typically 8.5\,hours, yielding a net integration time of 5 to 6 hours. 
The typical synthesized beam size (FWHM) was 1.5\,mas $\times$ 0.8\,mas 
with a position angle (P.A.) of $-30^{\circ}$. To obtain the 
positions of maser spots in S~Crt, the quasar J1147$-$0724 was 
simultaneously observed as the position reference. 
The J2000 \textit{a priori} coordinates of two sources are  
($\alpha$, $\delta$) $=$ 
(11\,h~52\,m~45.0981\,s, $-$07$^{\circ}$~35'~48.072'') for S~Crt, 
and (11\,h~47\,m~51.554035\,s, $-$07$^{\circ}$~24'~41.14109'')
for J1147$-$0724. 
J1147$-$0724 is classified as a ``candidate" source in the catalog 
of the International Celestial Reference Frame (ICRF), 
and its position errors are 270\,$\mu$as and 290\,$\mu$as in R.A.\ 
and Dec, respectively \citep{ma98}. 
The separation and P.A. of these sources are 
1.23$^{\circ}$ and $-80^{\circ}$. 

From a total of 17 observations, we used 12 observations in 
the present study. The observation in 2005 November was not 
used because the Ishigaki-jima antenna did not participate. 
Bad weather conditions in four observations held between August 
and October in 2006 resulted in much higher system 
noise temperatures ($\ge$500\,K) at some of the stations. 
The shorter integration times and unsolved phase fluctuations in 
these four observations degraded the image qualities of 
S~Crt in both self-calibration and phase reference analyses. 

The observation status is summarized in table~\ref{Observations}, 
which is organized as follows:
Column~(1)---The ID number of epoch. The numbers with an asterisk 
``*'' indicates that the observation was used in 
the estimation of the annual parallax. 
Column~(2)---Date of the observation. 
Column~(3)---The year and day of the year\,(DOY).   
In the observations in 2007/013, 2007/096, and 2007/130, 
the phase referenced images are scattered into several 
components, while the self-calibrated images 
resulted in a 
single bright component, clearly indicating that the atmospheric 
conditions on these days prohibited succesful 22\,GHz phase referencing. 
We did not use these three observations for the parallax estimation. 
In parallel with the VLBI program, we have monitored S~Crt 
with single-dish observations at Iriki station since 2003 September 
with the typical interval of one month (see also \cite{shi08}). 

A data recording rate of 1024\,Mbps was adopted with the 
VERA DIR2000 recording system, which yields the total 
receivable bandwidth of 256\,MHz with 2-bit digitization. 
The 256\,MHz bandwidth data of left-hand circular polarization 
were divided into 16 IF channels of 16\,MHz band width, one of 
which was used to receive the maser emission and the others 
were used to receive the continuum emission from J1147$-$0724. 
Cross-correlation was carried out with the Mitaka FX 
correlator~\citep{shi98} at the National Astronomical Observatory 
of Japan (NAOJ). 
In most observations, the IF channel 
assigned to the water maser was divided into 512 spectral 
channels, yielding a frequency spacing of 31.25\,kHz 
corresponding to a velocity resolution of 
0.42\,km\,s$^{-1}$. The observations in 2006/069 and 2006/129 provide 
two times higher frequency resolution than other observations, 
yielding a velocity resolution of 0.21\,km\,s$^{-1}$. 
For the data of J1147$-$0724, the IF channels were divided into 
64 spectral channels in all observations.

\subsection{Calibration and Imaging}
We used the Astronomical Imaging Package Software (AIPS) developed 
in the National Radio Astronomical Observatory in the data reduction. 
Amplitude calibration was achieved using the system 
noise temperatures and gains logged during the observations 
at each station. 
In the fringe search process of the reference source J1147$-$0724, 
we used the task {\sc fring} with a typical integration time of 
2\,minutes, with solutions of fringe phases, group delays, 
and delay rates obtained every 30\,seconds. 
Using the task {\sc tacop}, these solutions were transfered to  
the data of S~Crt in order to calibrate the visibility data. 
Phase and amplitude solutions obtained from self-calibration 
of the J1147$-$0724 were also transferred to the S~Crt data. 

Since the delay-tracking models used to estimate \textit{a priori} 
delays in the Mitaka FX correlator were not accurate enough for 
astrometry with VERA, we have applied better estimates 
calculated with the CALC3/MSOLV software package~\citep{jik05}. 
\citet{man91} gave a brief description of this package in 
the report to the International Earth Rotation Service (IERS). 
CALC3 has slight differences in the physical models than those 
adopted in CALC which was developed by the research group at 
NASA Goddard Space Flight Center (GSFC). 
Comparing the two \textit{a priori} values from Mitaka FX 
correlator and the CALC3/MSOLV package, we obtained the 
difference between the two estimates, then applied them to the 
visibility data using the tasks {\sc tbin} and {\sc clcal} in AIPS. 
We note that the zenith excess path lengths due to the 
wet atmosphere measured by the global positioning system 
(GPS) at each station are considered in the CALC3/MSOLV package. 

We estimated the residual atmospheric zenith delay offset 
using the method described in \citet{hon07}. 
This correction was applied to one of the four stations of VERA 
where the coherence of the phase-referenced image was maximized. 
Typically, the offsets were found within $\pm$3\,cm. 

The instrumental delay caused by the difference between two signal 
paths was estimated using an artificial noise source~\citep{kaw00}. 
These delays were then loaded into AIPS and applied in the 
same manner as the delay tracking model correction. 
For each observation, we fitted 
a two dimensional Gaussian model 
to the brightness distribution 
to find the position 
of the maser spot. Then we used these to derive the annual 
parallax and linear proper motions ($\mu_X$, $\mu_Y$). 

In the single-beam VLBI imaging of S~Crt, group delays solved with 
3C279 were applied, then fringe-fitting was done using the brightest 
emission showing the LSR velocity of 34.7\,km\,s$^{-1}$. 
Thus relative positions of all maser spots in self-calibrated 
images were determined with respect to this reference spot. 
For each velocity channel, position of the spot is defined 
as the brightness peak of the image. 
To estimate the relative motion of each spot, 
a linear least-squares analysis was applied to the 
spot seen at the same velocity channel during 
at least two continuous observations. 
The signal to noise ratio of 10 was adopted 
as the detection criterion in the self-calibrated images. 
The rms noise of the self-calibrated images was typically 
90\,mJy\,beam$^{-1}$. On the other hand, 
the rms noise of the phase-referenced images was typically 
700\,mJy\,beam$^{-1}$, about one order of magnitude larger 
than that of self-calibrated images. 
Due to limited phase coherence the phase referenced images 
have limited signal-to-noise (30 vs 307 in the self-calibrated 
images). 

\begin{table}
\caption{Observations}
\label{Observations}
\begin{center}
\begin{tabular}{clcc} 
\hline
ID    &       Date         & Year/DOY \\ \hline \hline 
1*    & 2005  October  19  & 2005/292 \\
2*    & 2005  December  3  & 2005/337 \\
3*    & 2006  January   5  & 2006/005 \\
4*    & 2006  February 11  & 2006/042 \\
5*    & 2006  March    10  & 2006/069 \\
6*    & 2006  May       9  & 2006/129 \\
7*    & 2006  November 13  & 2006/317 \\
8*    & 2006  December 11  & 2006/345 \\
9~    & 2007  January  13  & 2007/013 \\
10*   & 2007  February 21  & 2007/052 \\
11~   & 2007  April    06  & 2007/096 \\
12~   & 2007  May      10  & 2007/130 \\
\hline        
\end{tabular} 
\end{center}  
\end{table}

\section{Results}
\subsection{Annual parallax and distance} 
\begin{figure}
\begin{center}
 \includegraphics[width=80mm, angle=0]{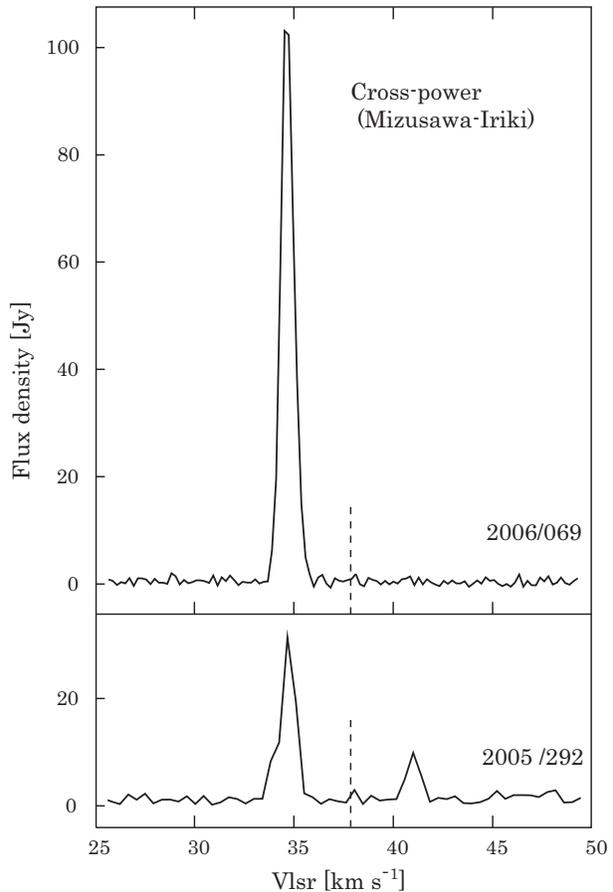}
 \caption{
Cross-power spectra of the S~Crt water masers observed with 
the Mizusawa-Iriki baseline on 2005/292 (top) and 2006/069 (bottom). 
The stellar velocity of 
37.85\,km\,s$^{-1}$ is indicated with the vertical dashed 
line in the spectra. 
The blue-shifted component with respect to the stellar velocity 
has been brighter than the red-shifted 
one in the majority of our observations. 
}
  \label{fig-spectr}
\end{center}
\end{figure}

In figure~\ref{fig-spectr}, cross-power spectra of S~Crt 
from 2005/019\,(upper) and 2006/069\,(lower) 
on the Mizusawa--Iriki baseline are presented. 
The time variation of the flux densities with the total 
power spectrum are presented in figure~\ref{lightcurves}(b) 
with an indication of each LSR velocity. 
The 34.7\,km\,s$^{-1}$ spot underwent a radio flare\,(e.g. \cite{shi05}). 
This flare started in February 2006 and reached a maximum of 
371\,Jy in March 31\,(2006/090), then it decreased to 60\,Jy in July. 
During the flare, this spot did not show structural change. 
The cross-power spectrum at the same time showed significantly 
weaker ($\simeq$30\%) flux density than that of the total 
power spectrum, 
indicating that some of the flared emission is 
resolved on the VERA baselines. 

\begin{figure}
\begin{center}
 \includegraphics[width=80mm, angle=0]{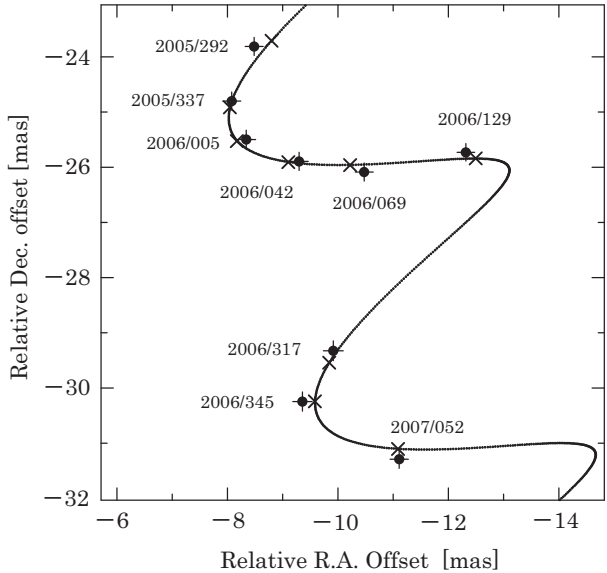} 
\caption{
Absolute positions of the reference maser spot in S~Crt. 
Filled circles indicate present results with error bars 
and crosses indicate predicted positions. 
Solid curve indicates the combined motion of the best-fit 
results of parallax and proper motion. 
The axes indicate the position offsets with respect to 
the phase tracking center.
}
  \label{fig-sky}
\end{center}
\end{figure}

In figure~\ref{fig-sky}, we present the positions of maser spot at the 
LSR velocity of 34.7\,km\,s$^{-1}$ relative to the phase tracking center. 
Throughout the present VLBI observations, this maser spot was  
bright enough to be detected on all baselines in the phase referencing 
analyses. The proper motion was clearly modulated by a parallax. 
Based on a least-squares fitting analysis, the parallax was determined to 
be 2.33$\pm$0.13\,mas which corresponds to a 
distance of 430$^{+25}_{-23}$\,pc. 
Here, we adopted position errors of each measurement that are obtained as the 
root sum squares of three error factors, and the details are given in section~4.1. 

In the estimation of the parallax, we adopt a very small number of 
assumptions: 
the maser spot is moving on a linear trajectory with respect 
to the star, i.e., there is no acceleration, and the reference source is 
fixed on the sky, i.e., no motions due to core shift or jet features. 
From the fitting results, the linear proper motions 
of the reference spot ($\mu_X$, $\mu_Y$) were obtained to be 
($\mu_X$, $\mu_Y$) $=$ 
($-1.56\pm0.22$\,mas\,yr$^{-1}$, $-5.16\pm0.22$\,mas\,yr$^{-1}$). 
This motion is the combination of the proper motion of S~Crt system 
and internal motion of the maser spot in the system. 
Considering the offset of the reference spot from the phase tracking center 
($\Delta\alpha$,$\Delta\delta$) $=$  ($-8.49$\,mas, $-23.81$\,mas), 
the J2000.0 absolute coordinates of this spot in 2005/292  
were obtained to be ($\alpha$,$\delta$) $=$ 
(11\,h~52\,m~44.96969\,s, $-$07\,$^{\circ}$~35'~48.0958''). 
This is the position as referenced to the position of J1147$-$0742 
and in relative offset to the original phase tracking centre.
The uncertainty of this position is estimated as 
$\sim$400\,$\mu$as based on the errors in our phase referencing 
analysis (detailed in section~4.1) and ICRF position of J1147$-$0724. 

The reference source J1147$-$0724 exhibited an unresolved structure.
The correlated amplitude as a function of ($u,v$) distance is 
flat and the upper limit of the source size is 0.8\,mas (FWHM), which 
is the minor axis of the synthesized beam in the present observation. 
In addition, we confirmed that the images of J1147$-$0724 
showed no distinctive change during the observations. 

\subsection{Maser distribution and internal motions}
\begin{figure*}
\begin{center}
\includegraphics[width=110mm, angle=0]{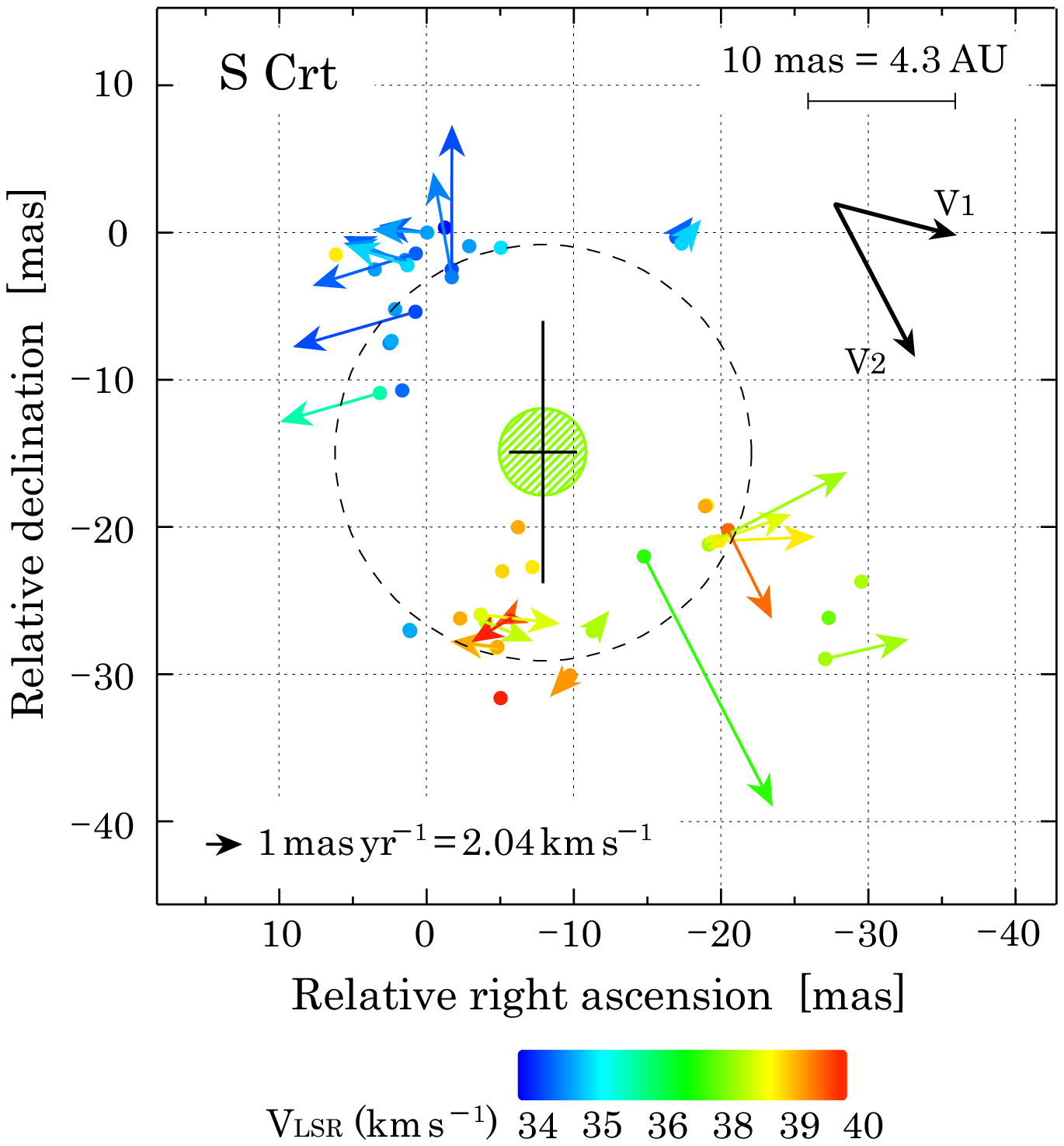}
\caption{
Angular distribution and internal motion vectors of maser spots in S~Crt.
The spot color indicates the LSR velocity (see color index at the bottom). 
The stellar velocity of S~Crt is 37.85\,km\,s$^{-1}$. 
The arrow at the bottom-left corner indicates a proper motion of 
1\,mas\,yr$^{-1}$, corresponding to 2.04\,km\,s$^{-1}$ at the distance of 
430\,pc. 
The shaded-circle indicates the size of the stellar photosphere. 
The dashed line represents a circular fit to the maser distribution, 
and has a radius of 14.9\,mas\,(see section~4.2 for details).  
The crosses indicate the stellar positions estimated from a different 
technique. 
The bold arrows $\mathbf{V}_1$ and $\mathbf{V}_2$ are the eigenvectors 
for the largest eigenvalue indicating the major axes of maser distribution. 
}
\label{fig-map}
\end{center}
\end{figure*}

In figure~\ref{fig-map}, we present the angular distribution 
and three-dimensional velocity field of maser spots in S~Crt 
covering a 60\,mas $\times$ 60\,mas region. 
The maser spot with the LSR velocity of 34.7\,km\,s$^{-1}$ 
is placed at the map origin. 
At the distance of 430\,pc, 
1\,mas corresponds to 0.43\,AU, and 1\,mas\,yr$^{-1}$ 
corresponds to a velocity of 2.04\,km\,s$^{-1}$. 
The color index in figure~\ref{fig-map} shows the 
LSR velocity range from 34.0 to 41.0\,km\,s$^{-1}$. 
The blue- and red-shifted components are 
separated into the northeast and southwest parts of the area. 
The relative motion of each spot ($v_{x}$, $v_{y}$) 
with respect to the reference spot are used to determine the 
average motion ($\bar{v_{x}}$, $\bar{v_{y}}$) and, hence we 
obtained ($\bar{v_{x}}$, $\bar{v_{y}}$) 
$=$ ($-$1.605\,mas\,yr$^{-1}$, $-$0.252\,mas\,yr$^{-1}$). 
Then, we subtracted ($\bar{v_{x}}$, $\bar{v_{y}}$) from 
($v_{x}$, $v_{y}$) to obtain the internal motions ($V_x$, $V_y$), 
that are presented with arrows in figure~\ref{fig-map}. 
We successfully detected the internal motions of 26 maser spots. 
The typical transverse speed was obtained to be 
2.72\,mas\,yr$^{-1}$, corresponding to 5.56\,km\,s$^{-1}$, 
by averaging the internal motions of all 26\,spots. 
The parameters of the motions are presented in table~\ref{spot} 
in the increasing order of LSR velocity. 

For the reference spot, we subtracted the internal motion from the 
proper motion ($\mu_X$, $\mu_Y$) and, thus, the proper motion 
of S~Crt system was estimated to be 
($-3.17\pm0.22$\,mas\,yr$^{-1}$, $-5.41\pm0.22$\,mas\,yr$^{-1}$). 
In the new HIPPARCOS catalog~\citep{van07}, the absolute proper 
motion of S~Crt was 
($-3.37\pm1.00$\,mas\,yr$^{-1}$, $-4.67\pm0.75$\,mas\,yr$^{-1}$), 
representing good consistency with our result within the errors. 

\begin{table*}
\caption{Parameters of the detected maser spots}
\label{spot}
\begin{center}
\begin{tabular}{ccccccccc} 
\hline
ID &$V_{\mathrm{LSR}}$&$X$ &$Y$  &$S$   & $V_x$  & $\sigma_{Vx}$ & $V_y$ & $\sigma_{Vy}$ \\ 
   &(km\,s$^{-1}$)&(mas) &(mas)      &(Jy\,beam$^{-1}$)&(mas\,yr$^{-1}$)&            & (mas\,yr$^{-1}$) &            \\ \hline\hline  
1  & 34.10 &  $-$1.25 &     0.34 &  2.3 &   \dots&  \dots&   \dots& \dots \\ 
2  & 34.53 &  $-$1.71 &  $-$2.50 &  2.7 & $-$0.01&  0.01 &    4.90& 2.58  \\
3  & 34.55 &     0.76 &  $-$5.39 &  4.4 &    4.15&  0.55 & $-$1.19& 0.06  \\
4  & 34.68 &     0.00 &     0.00 & 19.5 &    1.61&  \dots&    0.25& \dots \\
5  & 34.68 & $-$16.97 &  $-$0.34 &  4.6 & $-$0.55&  0.02 &    0.69& 0.09  \\
6  & 34.68 &  $-$2.85 &  $-$0.91 &  8.1 &   \dots&  \dots&   \dots& \dots \\
7  & 34.72 &     0.73 &  $-$1.42 &  1.8 &    3.50&  0.29 & $-$1.09& 0.45  \\
8  & 34.72 &     1.43 &  $-$2.00 &  1.9 &    2.10&  0.54 &    0.88& 0.14  \\
9  & 34.74 &     1.64 & $-$10.71 &  5.8 &   \dots&  \dots&   \dots& \dots \\
10 & 34.94 &  $-$1.72 &  $-$3.04 &  1.6 &    0.63&  0.50 &   3.56 & 0.37  \\
11 & 34.95 &     1.45 &  $-$1.86 & 35.4 &    2.08&  0.89 &   0.58 & 0.06  \\
12 & 34.97 &     2.51 &  $-$7.51 &  2.3 &   \dots&  \dots&   \dots& \dots \\
13 & 34.97 &     1.11 & $-$27.05 &  2.5 &   \dots&  \dots&   \dots& \dots \\
14 & 35.10 &  $-$0.03 &     0.01 &  8.4 &    1.85&  0.74 &   0.10 & 0.04  \\
15 & 35.10 &  $-$2.89 &  $-$0.92 &  4.9 &   \dots&  \dots&   \dots& \dots \\
16 & 35.16 &     3.52 &  $-$2.50 &  3.5 &   \dots&  \dots&   \dots& \dots \\
17 & 35.16 &     2.13 &  $-$5.20 &  3.2 &   \dots&  \dots&   \dots& \dots \\
18 & 35.18 &     2.38 &  $-$7.37 &  2.1 &   \dots&  \dots&   \dots& \dots \\
19 & 35.18 &     1.14 & $-$27.01 &  2.1 &   \dots&  \dots&   \dots& \dots \\
20 & 35.52 &     1.30 &  $-$2.21 &  1.5 &    2.05&  0.90 &   0.68 & 0.36  \\
21 & 35.52 & $-$17.32 &  $-$0.74 &  1.3 & $-$0.66&  0.06 &   0.80 & 0.11  \\
22 & 35.54 &  $-$5.04 &  $-$1.04 &  0.9 &   \dots&  \dots&   \dots& \dots \\
23 & 36.45 &     3.17 & $-$10.90 &  1.3 &    3.39&  0.52 & $-$0.98& 0.49  \\
24 & 37.89 & $-$14.77 & $-$22.00 &  0.8 & $-$4.37&  0    & $-$8.49& 0     \\
25 & 38.32 & $-$27.32 & $-$26.17 &  1.3 &   \dots&  \dots&   \dots& \dots \\
26 & 38.73 & $-$27.06 & $-$28.97 &  1.0 & $-$2.83&  0.15 &   0.67 & 0.45  \\
27 & 38.73 & $-$19.16 & $-$21.19 &  0.9 & $-$4.70&  0    &   2.46 & 0     \\
28 & 38.74 & $-$29.53 & $-$23.72 &  0.9 &   \dots&  \dots&   \dots& \dots \\
29 & 38.74 & $-$11.30 & $-$27.07 &  1.0 & $-$0.55&  0.02 &   0.69 & 0.09  \\
30 & 38.91 &  $-$4.01 & $-$26.46 &  1.1 & $-$1.60&  0.43 & $-$0.65& 0.29  \\
31 & 39.15 &  $-$3.67 & $-$25.96 &  0.8 & $-$2.69&  0.30 & $-$0.30& 0.18  \\
32 & 39.15 & $-$19.40 & $-$21.04 &  2.2 & $-$2.67&  0.37 &   0.95 & 0.44  \\
33 & 39.57 & $-$19.87 & $-$20.95 &  3.4 & $-$3.26&  0    &   0.14& 0      \\
34 & 39.57 &  $-$7.18 & $-$22.73 &  2.2 &   \dots&  \dots&   \dots& \dots \\
35 & 39.57 & $-$19.02 & $-$18.52 &  2.0 &   \dots&  \dots&   \dots& \dots \\
36 & 39.59 &     6.15 &  $-$1.49 &  1.8 &   \dots&  \dots&   \dots& \dots \\
37 & 39.61 &  $-$4.73 & $-$28.23 &  0.7 &    1.49&  0    &    0.23& 0     \\
38 & 39.74 &  $-$5.14 & $-$23.02 &  1.0 &   \dots&  \dots&   \dots& \dots \\
39 & 40.00 & $-$18.91 & $-$18.58 &  1.2 &   \dots&  \dots&   \dots& \dots \\
40 & 40.01 &  $-$6.22 & $-$20.02 &  4.7 &   \dots&  \dots&   \dots& \dots \\
41 & 40.03 &  $-$4.81 & $-$28.14 &  1.0 &    1.59&  0    &    0.18& 0     \\
42 & 40.03 &  $-$2.29 & $-$26.22 &  2.0 &   \dots&  \dots&   \dots& \dots \\
43 & 40.16 &  $-$9.75 & $-$30.09 &  1.6 &    0.68&  0.38 & $-$0.72& 0.06  \\
44 & 40.42 & $-$20.47 & $-$20.21 &  1.1 & $-$1.49&  0.23 & $-$3.01& 0.51  \\
45 & 40.58 &  $-$5.73 & $-$26.12 &  2.6 &    0.54&  0.05 & $-$0.26& 0.17  \\
46 & 41.00 &  $-$5.03 & $-$31.63 &  1.1 &   \dots&  \dots&   \dots& \dots \\
47 & 41.00 &  $-$5.70 & $-$26.20 &  5.4 &    1.32&  1.27 & $-$0.81& 0.42  \\
\hline
\multicolumn{9}{@{}l@{}}{\hbox to 0pt{\parbox{140mm}{\footnotesize
\smallskip
Column~(1)---Component ID.
Column~(2)---LSR velocity in km\,s$^{-1}$. 
Column~(3)---offset positions in R.A. relative to the original phase center. 
Column~(4)---offset positions in Dec. relative to the original phase center.
Column~(5)---brightness of the spot at the first detection in Jy\,beam$^{-1}$. 
Column~(6)---best fit linear motion in R.A. in mas\,yr$^{-1}$.
Column~(7)---standard errors of the motions in R.A. 
In case the number of successful detections were restricted to be only twice, 
we adopted the standard errors $\sigma_X$ and $\sigma_Y$ for the spot as zero. 
Column~(8)---best fit linear motion in Dec in mas\,yr$^{-1}$.
Column~(9)---standard errors of the motions in Dec. 
}\hss}} 
\end{tabular} 
\end{center}  
\end{table*}   

\section{Discussion} 
\subsection{Errors in the positions}
Since it is difficult to identify contributions to positional errors
in individual VLBI observations
~\citep{kur05, hac06, hon07, sat07, hir07, ima07}, 
we applied errors in each observation 
by evaluating following three error sources: 
(1)\,a zenith atmospheric delay offset at each station, 
(2)\,station position errors, and 
(3)\,image quality.  
The best estimates of the zenith atmospheric delay offsets 
fell within the range of $\pm$3\,cm.
Errors in the \textit{a priori} station positions can result in 
additional phase residuals.
Therefore, we assumed the offset of 3\,cm even after the 
atmospheric offset calibration was done. 

For a given separation of zenith angles between two sources 
$\Delta Z$, the difference of signal path lengths 
$\Delta l$ between the two directions caused by the zenith 
atmospheric delay residual $l_0$ can be estimated as follows, 
\begin{eqnarray}
\Delta l &= l_0\,\Delta Z\,\frac{d}{d Z} \left(\frac{1}{\cos Z}\right), 
\label{deltaSecZ}
\end{eqnarray}
where $Z$ is the mean zenith angle of the observed sources. 
We adopted 1.23$^{\circ}$ as the approximation of $\Delta Z$. 
During our observations, the elevation angles higher than 
30$^{\circ}$ were dominant. We assumed $Z=50^{\circ}$ as 
the zenith angle for estimating errors of the worst case, and 
obtained $\Delta l =$ 0.12\,cm. 
Using a root sum squares of major and minor axes of the 
synthesized beam\,($\theta_b$ = 1.7\,mas), 
we estimated a positional error of $\sigma_Z =$ 156\,$\mu$as 
from the following expression,  
\begin{eqnarray}
\sigma_Z = \theta_b \times \frac{\Delta l}{\lambda_{H_2O}}, 
\end{eqnarray}
where $\lambda_{H_2O}$ is the wavelength corresponding to  
the rest frequency of water masers. 
Since the position accuracy of antennas are determined to be 
$\sim$3\,mm based on geodetic observations at S and X bands, 
we estimated the positional error\,($\sigma_D$) attributed to 
the baseline errors to be 5\,$\mu$as using the following 
expression, 
\begin{eqnarray}
\sigma_D = Sin\theta_{SA} \times \frac{3\,mm}{\lambda_{H_2O}}, 
\end{eqnarray}
where $\theta_{SA} = 1.23^{\circ}$ is the separation 
angle of the source pair. 
In addition, we estimated the measurement errors
\,($\sigma_I = \theta_b/SNR$) which depends on the SNRs of 
the phase-referenced images. 
Finally, we obtained the position errors of each phase-referenced 
image by adding up quadratically the error factors, 
$\sigma_Z, \sigma_D$, and $\sigma_I$. 
The error bars in figure~3 represent the errors in each observation, 
and the averaged value of the errors is 167\,$\mu$as. 

\subsection{Maser Morphology}  
Figure~\ref{fig-map} shows the internal motions of the masers pots 
around S~Crt found in the present observations. 
Since the maser spots show a bimodal distribution about radial 
velocities and positions, we extract the kinematic essentials. 
The analytic tools based on the diagonalization of the velocity 
variance-covariance matrix (VVCM)~\citep{blo00} and 
position variance-covariance matrix are used. 

The elements of the VVCM matrix are presented as 
\begin{eqnarray}
\sigma_{jk} = 
\frac{1}{N-1} \sum\limits_{i=1}^{N} (V_{j,i}-\bar{V_{j}})(V_{k,i}-\bar{V_{k}}), 
\label{VVCM}
\end{eqnarray}
where the diagonal elements are the velocity dispersion. 
Here $j$ and $k$ denote three orthogonal space axes 
(e.g., R.A., Dec., and radial coordinate $z$), 
$i$ denotes the $i$th maser spot in a collection totalling N($=$26), 
and the bar indicates averaging over maser spots. 

The vector $\mathbf{V}_1$~(figure~\ref{fig-map}), which is a two-dimensional 
projections of the eigenvector for the largest eigenvalue, is the major 
principal axis of the VVCM. 
This axis of maximum internal velocity dispersion is the outflow axis.  
The axis of $\mathbf{V}_1$ lies at a P.A. of $\sim255^{\circ}$, and also 
make an inclination angle $\sim43^{\circ}$ with the plane of the sky, 
with the southwest lobe directed into the page (away from the observer).
The VVCM was initially evaluated in a coordinate system with axes 
R.A., Dec., and radial coordinate z increasing toward the source, 
and then it was diagonalized. 

We also considered the two-dimensional variance-covariance 
matrix to quantify the angular distribution of the maser spots. 
Here, the number of maser spots is 26, same as the case of VVCM. 
In the diagonalization of this matrix, one eigenvalue which 
is larger than the other by a factor of 3.1 was obtained. 
The corresponding eigenvector $\mathbf{V}_2$ presented in 
figure~\ref{fig-map} indicate the spatial elongation of the 
maser distribution. 
The position angle of this axis is found to be $\sim208^{\circ}$. 
Although maser velocity and position axes show a discrepancy 
of $\sim47^{\circ}$, the directions of these model-independent 
axes indicate the bipolar outflow with the major axis along northeast-southwest.

This dynamical property is not revealed by previous 
observation \citep{bow94}. 

We estimate the stellar position from the present result by 
using the kinematic information. 
Since the motion of each maser spot are powered by the forming star, 
its position can be estimated as an origin of these motions. 
Here, we take one of the simplest kinematic models, which assumes 
a linear motion with a constant velocities from a single origin. 
This technique is essentially same as that used in \citet{ima00} 
and \citet{hon07}. 
Based on this modeling, we obtained the origin of the maser 
motions which most effectively explain the observed positions 
and velocities.  
The stellar position from this kinematic analysis, the kinematic center 
was obtained to be 
($X$, $Y$) $=$ ($-$7.9$\pm$1.8\,mas, $-$14.9$\pm$13.3\,mas), 
which is indicated with the cross with error bars in figure~\ref{fig-map}. 
We used the first order moment with respect to the determined stellar 
position in order to find a typical distance of maser spots. 
The obtained distance was 14.2$\pm$3.5\,mas, and presented with dashed 
line in figure~\ref{fig-map}. 
At the distance of 430\,pc, the corresponding linear length is 6.11\,AU. 
Since we estimated the inclination angle of the outflow axis, 
this projected length can be converted to the linear radius of 8.96\,AU. 

Nineteen years ago, the angular distribution of the masers 
was firstly observed with the NRAO Very Large Array in its 
A configuration \citep{bow94}. 
The masers were uniformly distributed over a 40\,mas$
\times$40\,mas region with the LSR velocities ranging 
evenly from 33.4 to 44.2\,km\,s$^{-1}$ (figure~13 of \cite{bow94}). 
In their study, the blue-(red-)shifted spots were found 
in the northeast(southwest) of the maser region. 
In addition, masers with intermediate velocities were found 
in the central region. 
Although the extent of the distribution is consistent 
with our result, we did not detect the masers in central region. 
In figure~\ref{fig-map}, the maser spots are mainly found 
at the peripheral region of the distribution. 
During the period of 2006.3 -- 2007.0, we could not find the emission 
with intermediate velocities(see figure~\ref{fig-spectr}(b)), and it 
can explain the the absence of intermediate velocity maser spots . 
On the other hand, at the beginning and end of our VLBI observations, 
the masers with intermediate velocities are found in total power 
spectrum, and they distributed in the peripheral region. 
As \citet{bow94} did not precisely mention the flux densities and sizes 
of each spot, it is difficult to specify the reason for our non-detection 
of the masers in central region. 
However, we can speculate as to the reason; 
(1)\,Our image sensitivities were not high enough to detect them. 
The rms noise 38\,mJy\,beam$^{-1}$ of \citet{bow94} was lower than 
that of the present VLBI observations, 90\,mJy\,beam$^{-1}$. 
(2)\,The masers in central region have faded  
before the present VLBI observations. 
(3)\,The masers were extended and diffuse, therefore they were resolved 
out with the synthesized beam of VERA. 
For example, we find the effect of resolve-out (e.g. $\leq$30\,\% 
in 2006 March) from the comparison between cross-power spectrum 
(figure~\ref{fig-spectr}) and total-power spectrum (figure~\ref{lightcurves}(b)). 

\subsection{Stellar diameter and temperature}  
We extract the stellar properties of S~Crt using the distance. 
\citet{ari99} determined the temperature of photosphere 
($T_{\mathrm{BB}}$) in S~Crt to be 3097$\pm$100\,K by fitting 
two blackbodies to its infrared spectrum. 
Using this temperature, we estimated an acceptable temperature 
range of 2600 -- 3500\,K. 
Here we assume that the light variation of 0.8\,mag 
(figure~\ref{lightcurves}(a)) is same in the infrared and 
caused only by the temperature variation. 
\citet{ari99} reported that Mira variables show $T_{\mathrm{BB}} =$ 
2418 -- 2902\,K and SRVs shows $T_{\mathrm{BB}} =$ 3005 -- 3500\,K. 
The estimated temperature range of S~Crt is consistent with their 
finding. 

It is worthwhile estimating the stellar radius of S~Crt, 
since there is no direct measurement. 
Using the apparent magnitudes in infrared\,(J, H, and K), 
$T_{\mathrm{BB}}$\,(3097\,K) and the distance\,(430\,pc), 
we estimated the radius of S~Crt photosphere ($R_{*}$) 
to be $1.81\pm0.14\times10^{13}$\,cm\,(260$\pm$20\,$R_{\odot}$). 
This is presented with a shaded-circle in figure~\ref{fig-map}, 
and the center of circle is placed on the kinematic center. 
Using this radius, we estimated an acceptable radius range of 
213 -- 309$R_{\odot}$, under the assumption that the light 
variation is caused, now, only by the radius variation. 
In the studies of \citet{han95} and \citet{van97}, $R_{*}$s of dozen 
of Mira variables are presented, and most $R_{*}$/$R_{\odot}$ is 
larger than 300. 
The photospheric radius of S~Crt is close to 
the lower limits to those of Mira variables in the literature. 
Since the estimates of $R_{*}$ and $T_{\mathrm{BB}}$ are now available, 
the luminosity ($L_{*}$) of S~Crt is estimated to be 
2.29$\times$10$^{30}$\,[W], resulting in the ratio $L_*$/$L_{\odot}$ 
of 5970. 

A similar analysis of data for a larger number of Mira and 
SRVs would be very useful for a better 
understanding of the difference and similarities between 
Mira variables and SRVs. 

\section{Summary} 
We have used VLBI monitoring observations with VERA to measure the 
absolute proper motions of a maser spot in S~Crt with respect 
to the quasar J1147$-$0724. We obtained an annual parallax of 
2.33$\pm$0.13\,mas corresponding to a distance of 430$^{+25}_{-23}$\,pc. 

Nearly two years astrometric observations revealed the internal 
motions of 26 maser spots for the first time. 
Taking into account the internal motions of the reference maser spot, 
the absolute proper motion of S~Crt system was obtained to be 
($-3.17\pm0.22$\,mas\,yr$^{-1}$, $-5.41\pm0.22$\,mas\,yr$^{-1}$), 
which is consistent with the result in the new HIPPARCOS catalog. 

The three-dimensional velocity field of the maser spots in S~Crt was 
also detected, clearly outlining a bipolar outflow with the major 
axis along northeast-southwest. The inclination of the axis 
with respect to the sky plane was $\sim43^{\circ}$. 
The angular size of the maser distribution can be converted to the 
linear maser distribution radius of 8.96\,AU. 

Using the photospheric temperature of 3097\,K~\citep{ari99}, the $K$ 
band apparent magnitude $m_{K}=$0.79, and the distance of 430\,pc, 
the linear radius of S~Crt photosphere ($R_{*}$) was estimated 
to be $1.81\pm0.14\times10^{13}$\,cm\,(260$\pm$20\,$R_{\odot}$). 
This radius is comparable with the lower limits to those of Mira 
variables in literatures. 

{}

\end{document}